\documentclass[a4paper,10pt]{article}
\usepackage[utf8]{inputenc}
\usepackage{graphicx}
\usepackage{physics}
\usepackage{authblk}
\usepackage{hyperref}
\newcommand*{\email}[1]{%
    \normalsize\href{mailto:#1}{#1}\par
    }
\title{Design and Optimization of Ellipsoid Scatterer-based Metasurfaces via the Inverse T-Matrix Method}
\author{Maksym Zhelyeznyakov$^{1,\#}$, Alan Zhan$^{2,\#}$, Arka Majumdar$^{1,2,*}$}
\affil{Department of Electrical and Computer Engineering, University of Washington, Seattle, Washington 98195, USA}
\affil{Department of Physics, University of Washington, Seattle, Washington 98195, USA}
\affil{$^\#$Contributed equally}
\affil{$^*$Corresponding author \email{arka@uw.edu}}

\begin{document}

\maketitle

\begin{abstract}
Large-area metasurfaces composed of discrete wavelength-scale scatterers present an extremely large number of degrees of freedom to engineer an optical element. These degrees of freedom provide tremendous design flexibility, and a central challenge in metasurface design is how to optimally leverage these degrees of freedom towards a desired optical function. Inverse design can be used to explore non-intuitive design space for metasurfaces. We report an inverse design method exploiting T-Matrix scattering of ellipsoidal scatterer based metasurfaces. Multi-functional, polarization multiplexed metasurfaces were designed using this approach. Finally, we apply this method to optimize the efficiency of an existing high numerical aperture (0.83) metalens design, and report an increase in efficiency from 26\% to 32\%.
\end{abstract}

\section{Introduction}
The design of optical elements made of quasi-periodic arrays of sub-wavelength scatterers, known as metasurfaces, is a promising area of research. The minituarization of existing optical elements such as lenses \cite{Chen2018, Zhan:16, Arbabi2015}, freeform optics \cite{Zhan2017}, and retroreflectors \cite{Arbabi2017} has already been shown using metasurfaces. Furthermore, multi-functional optical elements \cite{ArbabiPhase,AngleKamali} and new point spread function engineering methods \cite{Colburneaar2114,Colburn2019} have been demonstrated using metasurfaces. Until recently however, these metasurfaces have generally been designed intuitively, termed here as forward design. Libraries of complex transfer coefficients of individual scatterers are pre-computed, and arranged in a periodic lattice to approximate a desired phase profile. The properties of these scatterers are computed with periodic boundary conditions and the metasurfaces are designed under the "local phase approximation": the scattering in any small region is taken to be the same as the scattering from a periodic surface \cite{Hsu:17}. This approximation neglects inter-scatterer coupling  which is significant for metasurfaces composed of scatterers with rapidly varying geometries or with low refractive index \cite{Bayati:19}. Moreover, it is not always possible to know the phase-profile a priori, and in these cases forward design methods cannot be used.

Inverse design methods use a figure of merit (FOM) written in terms of adjustable geometric scatterer parameters to iteratively optimize the scatterers of a metasurface to implement a desired functionality. The inverse design approach starts with some arbitrary initial scatterer configuration. Then the electric field scattered off the device, the FOM, and the gradient of the FOM with respect to the scatterer design parameters are computed. The scatterer geometries are then iteratively updated in the direction that optimizes the FOM. Thus, inverse design methods offer a clear path to create optical elements with unintuitive phase functions. Different optimization methods such as particle swarm optimization \cite{Wang2014}, genetic algorithm methods \cite{Aieta1342,Egorov:17,Donelli:13}, and gradient based methods \cite{Zhan:18,Hansen:15,Piggott2017,Shen2015,Piggott2015,doi:10.1002/lpor.201000014,Pestourie:18,Lin:19,Fan:Inverse} have already been applied to design both integrated nanophotonic elements and free space metasurface optics. One specific direction is to exploit Mie scattering of spherical scatterers to perform the inverse design \cite{Zhan:18}. This approach allows large-area metasurface design without relying on the local phase approximation, and thus accurately models the inter-scatterer coupling. Currently, this approach is restricted to spherical scatterers, for which the radii are the only free parameters available when designing the metasurface. Additionally, we did not find a radius range over which these spherical scatterers smoothly span a $0-2\pi$ phase shift without suffering considerable optical losses, a common requirement when designing metaphotonic structures. Another drawback of gradient based photonic inverse design methods is the non-convexity of the FOM. The initial conditions for designing photonic devices critically affect the final performance, yielding in vastly differing designs.

 In this work, we present an inverse-design and optimization method for large-area ($\sim$ 40$\lambda$ in diameter) metasurface based on transition matrix scattering theory, an extension of Generalized Multi-sphere Mie Theory (GMMT). Specifically, we extended a previously reported inverse design method \cite{Zhan:18} to ellipsoidal scatterers. We first show the feasibility of using this method for inverse-designing single wavelength metasurfaces lenses. We then demonstrate the effectiveness of this method for designing non-intuitive devices without a known phase function by optimizing a polarization multiplexed lens that switches the location of the focal spot of a lens based on the polarization of the incident light. Finally we demonstrate the efficacy of inverse design techniques for optimization, by improving the efficiency of a high numerical aperture lens via optimization, starting with a forward designed metalens as the initial condition.

\section{T-Matrix formalism}
\begin{figure}[hbt!]
\centering\includegraphics[scale=0.4]{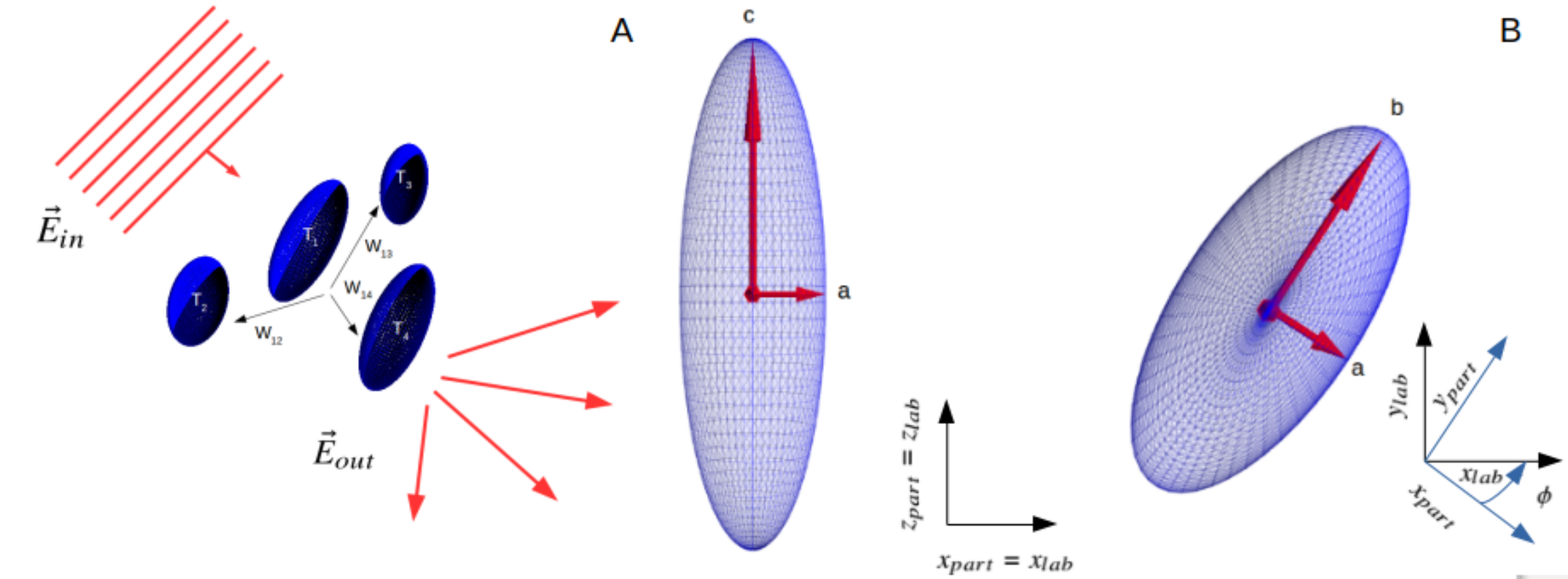}
\caption{\textbf{A.} Mie scattering schematic. Light is incident onto the set of ellipsoidal scatterers. Each scatterer has an associated T-Matrix. The incident field onto each scatterer is described by the incident field $E_{in}$ and the scattered fields from all other scatterers. The inter-particle coupling is represented by the matrix $W_{n''n'}^{i''i'}$ which describes the coupling between spheres $i'$ and $i''$. \textbf{B.} Design parameters for ellipsoidal scatterers. The semi-major axes are taken to always be aligned with the particle frame: semi-major axis $a$ is aligned with the $x_{part}$ axis, $b$ with the $y_{part}$ axis, and $c$ with the $z_{part}$ axis. The roation $\phi$ is about the z-axis, with the counterclockwise direction defined as a positive rotation.}
\label{fig:fig1}
\end{figure}

We adopt an adjoint based optimization approach by using the T-Matrix Method (TMM) for ellipsoidal scatterer geometries. Since a rigorous treatment of TMM for solving electromagnetic scattering from ensembles of scatterers can be found elsewhere \cite{Mackowski:96, Xu:95, EGEL2017103}, we provide give a brief overview of this method. The TMM formalism allows for a faster and less memory intensive forward simulation compared to direct methods of solving Maxwell's equations such as finite difference methods, at the cost of a restricted scatterer geometry.
	In the case of a single scatterer $S_i$, the net electric field can be written as a sum of the incident and scattered fields as $E(\vec{r}) = E_{in}^i(\vec{r}) + E_{scat}^i(\vec{r})$ where $E_{in}$ and $E_{scat}$ can be written as series expansions of the incident and scattered fields in the spherical vector wave function (SVWF) basis:
\begin{equation}
E_{in}^i = \sum_n a_{n}^i \psi_n^{(1)}(\vec{r}-\vec{r_i})
\end{equation}
\begin{equation}
E_{scat}^i = \sum_n b_{n}^i \psi_n^{(3)}(\vec{r}-\vec{r_i})
\end{equation}
Where $\psi_n$ are the SVWF of different orders, $a_n^i$ and $b_n^i$ are the coefficients of the incoming and scattered field from the $i^{th}$ scatterer respectively. $n$ is a multi-pole expansion index, that includes the orbital index $l$, azimuthal index $m$, and polarization index $p$. For the case of multiple spherical scatterers, the field can be written as 
\begin{equation}
E_{in}^i(\vec{r}) = E_{in}(\vec{r})+\sum_{i'\neq i}E_{scat}^{i'}(\vec{r})
\end{equation}
Where $E_{in}^{i}(\vec{r})$ is the incident field on the $i^{th}$ scatterer, $E_{in}(\vec{r})$ is the original incident field, and $E_{scat}^{i'}(\vec{r})$ is the scattered field from the $i'$th scatterer. The coefficients $b_n^i$ and $a_n^i$ for a single scatterer are related by the T-Matrix:
\begin{equation}
b_n^i = T_{nn'}^{ii'} a_n^i
\end{equation}
In the case of multiple scatterers we need to solve a system of coupled linear equations for $b_n^i$:
\begin{equation}
M_{nn'}^{ii'} b_{n'}^{i'} = T_{nn'}^{ii'}a_{in,n'}^{i'}\
\end{equation}
\begin{equation}
M_{nn'}^{ii'} = \delta_{ii'}\delta_{nn'} - T_{nn''}^{ii''}W_{n''n'}^{i''i'}
\end{equation}
with $a_{in,n'}^{i'}$ representing the coefficients that correspond to the incident field, and $W_{n''n'}^{i''i'}$ is the coupling matrix that relates the scattered field of the $i'$th scatterer to that of the $i''$th's incident field. The forward problem is solved via CELES, a CUDA-accelerated matlab package \cite{EGEL2017103} that allows for the simulation of scattering from large aggregates of spherical scatterers, with modifications to the T-Matrix definitions in order to simulate ellipsoidal scatterers. Fig. \ref{fig:fig1}A shows a schematic of how the forward problem is solved.

\section{Adjoint optimization}

Our design process begins with a set of scatterer locations and geometric properties of each individual ellipsoid as initial conditions. During the optimization process, each individual scatterer geometry is iteratively modified. To calculate the gradient, we use the adjoint method\cite{Hansen:15, Lalau-Keraly:13, Piggott2015, Piggott2017, Zhan:18}. In our previous work \cite{Zhan:18}, we calculated the general gradient of a FOM with respect to individual sphere radii \textbf{R}. Using a similar approach, we calculate the gradient of the FOM with respect to the free parameters of an elliptical scatterer. Fig. \ref{fig:fig1} B shows the free parameters of an ellipsoidal scatterer. Given a set of design parameters $\{\textbf{P}\}$, we can write a FOM $f(\textbf{b(P)},\textbf{P})$, where \textbf{b} is the vector containing coefficients $b_n^i$. We want to calculate the FOM with respect to parameters $\textbf{P}$. The procedure of calculating the gradient with respect to the free parameters of the ellipsoidal scatterer is identical to that of a spherical scatterer, so we refer to Eq. 13 of Ref. \cite{Zhan:18} and write
\begin{equation}
\frac{\partial f}{\partial P_j} = 2 Re \bigg\{(\lambda_n^i)^T \left(\frac{\partial T_{nn'}^{ii'}}{\partial P_j} a_{in,n'}^{i'}+\frac{\partial T^{ii''}_{nn''}}{\partial P_j} W^{i''i'}_{n''n'}b_{n'}^{i'}\right)\bigg\}
\end{equation}
Here, $\frac{\partial f }{\partial P}$ refers to the gradient of the FOM with respect to one of the principal semi axes of the ellipsoid ($a,b,$ or $c$), or its azimuth rotation $\phi$. $\lambda_n^i$ is the ``adjoint'' term given by $(\lambda_n^i)^T = \frac{\partial f}{\partial b_n^i}^T$. The terms $\frac{\partial T}{\partial P}$ refer to the gradients of the T-matrices with respect to the design parameters of the ellipsoids. The derivation of the T-Matrix gradients with respect to $a,b,c,\phi$ is detailed in the appendix.

\begin{figure}[hbt!]
\centering\includegraphics[scale=0.4]{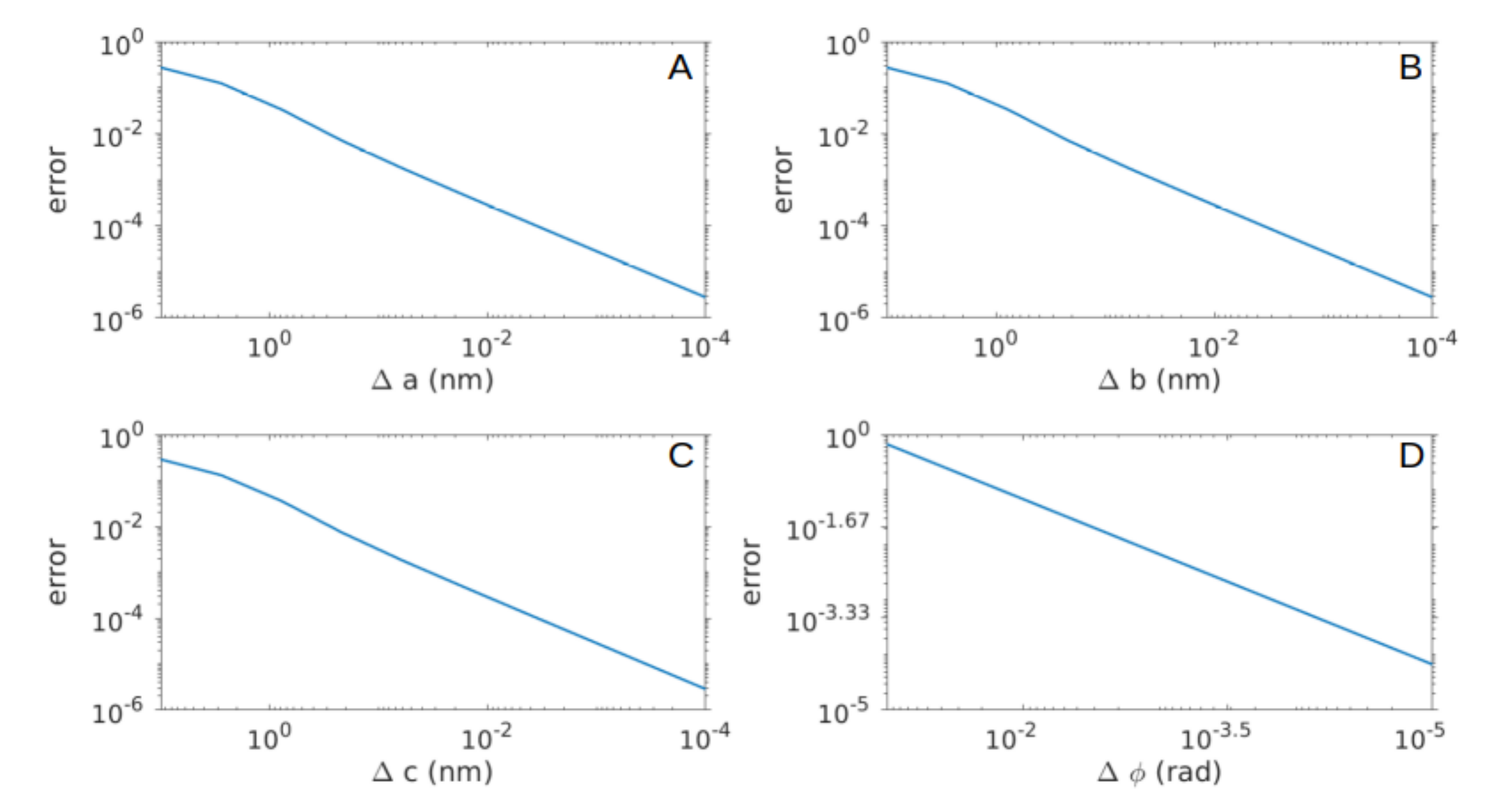}
\caption{Verificaton of the analytical T-Matrix derivaties. \textbf{A} shows the error between the analytical T-Matrix derivative and the numerical derivative with respect to semi-major axis $a$, \textbf{B} with respect to $b$, \textbf{C} with respect to $c$. Fig. \ref{fig:derivatives}D shows the T-Matrix derivative with respect to the azimuthal rotation of the ellipsoid $\phi$. As the step size of the numerical approximation to the derivative gets smaller, the mean error between numerical and analytical derivative gets closer to 0, which implies that the analytical derivatives are valid.}
\label{fig:derivatives}
\end{figure}

Since, to the best of our knowledge, this is the first time these gradients have been calculated, we first numerically verify their validity. We denote the T-Matrix of an ellipsoidal scatterer $S$ as a function of its geometry
\begin{equation}
T(S(P)) = T(S(a,b,c,\phi)) = T(a,b,c,\phi)
\end{equation}
We denote the numerical derivative of the T-Matrix as $\partial_P T^{N}$ with respect to some design parameter $P \in \{a,b,c,\phi\}$, and the analytical derivative as $\partial_P T^A$. We approximate the numerical derivative as 
\begin{equation}
\partial_P T^N = \frac{T(P+\Delta P) - T(P)}{\Delta P} + O(\Delta P^2)
\end{equation}
To validate the accuracy of the analytical derivative, we  create a set of 216 ellipsoids with geometries corresponding to permutations of $a,b,c$ between $50nm$ and $300nm$ in steps of $50nm$. We compute the T-matrices and their analytical derivatives with respect to each of the design parameters for each individual ellipsoid. We then compute the numerical derivative of each T-Matrix by using Eq. 9. Finally, we define the mean error of the derivatives by 
\begin{equation}
error = mean\left(\sum_i \sum_j \bigg|\partial_P T^N_{i,j} - \partial_P T^A_{i,j}\bigg|\right)
\end{equation}
where indices $i,j$ are the individual elements of the T-matix. We vary the step sizes for $\Delta a, \Delta b,$ and $\Delta c$ from $10$ to $10^{-4}$ nm and for $\Delta \phi$ from $10^{-1}$ to $10^{-5}$ radians. We show the plot of the mean error vs the step size of the numerical gradient in	Fig. \ref{fig:derivatives}. As the step size is reduced, the numerical derivative converges closer and closer to the analytical derivative, as expected.

\section{Inverse design of optical elements}
Using the aforementioned TMM formalism and adjoint optimization method, we present the design and optimization of two optical elements: a lens with numerical aperture (NA) of $\sim 0.83$, and a lens that switches focal lengths based on the polarization of light incident onto the lens. To design these devices, we must specify a FOM that encompasses its performance. In both cases, the lenses were designed for $915 nm$ incident wavelength, with scatterers having refractive index of 3.56 surrounded by vacuum. Each device was designed by starting off with identical ellipsoids, and minimizing a specified FOM.

\subsection{High numerical aperture lens}

\begin{figure}
\center\includegraphics[scale=0.35]{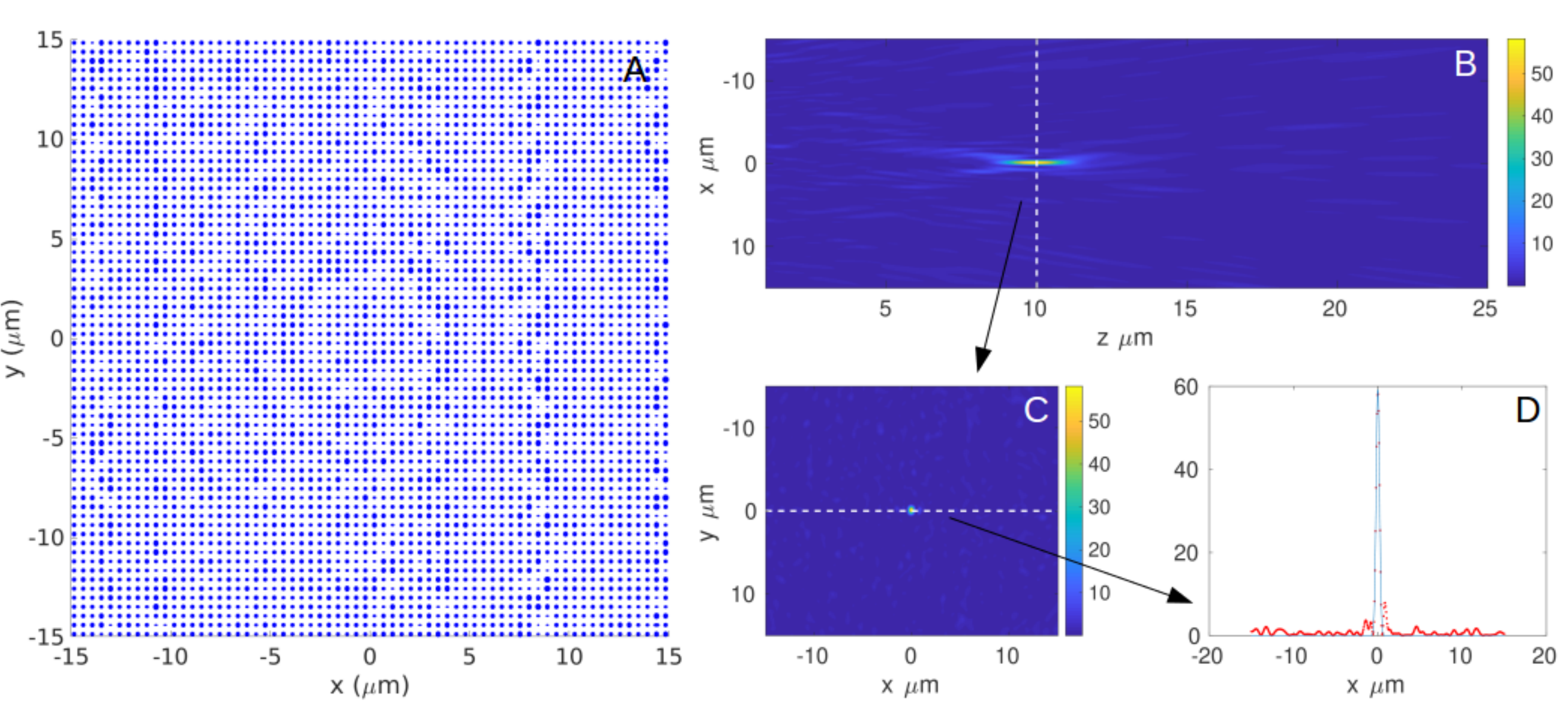}
\caption{\textbf{A} Final distribution of scatterers with periodicity 450 nm for the inverse designed lens. semi-major axes a and b are allowed to range between 40 and 150 nm. Semi-major axis c is allowed to range between 40 and 300 nm. \textbf{B} the field cross-section in the x-z plane at $y=0 \mu m$, \textbf{C} the cross-section in the x-y plane at $z=10\mu m$. \textbf{D} shows the Gaussian fit to the field at the focal spot $z=10 \mu m$ along $x=0$. In order to calculate the lens efficiency, the full-width at half-maximum (FWHM) was calculated for the fitted Gaussians. The integral of the field intensity around the disk $d=3 \times FWHM$ about the center of the focal spot was calculated, and then divided by the total incident field intensity. The units of all plots are arbitrary light intensity units. The efficiency of the inverse designed lens was calculated to be $3.38\%$.}
\label{fig:inverse_fields}
\end{figure}

Fig. \ref{fig:inverse_fields}A shows the distribution of ellipsoidal scatterers for a high NA lens with a diameter of 30 $\mu m$ and a focal length of 10 $\mu m$. The FOM that encompasses lens functionality can be written as:

\begin{equation}
f(b(P),P) = (I^T-I^A(x,y,z=F))^2
\end{equation}

Here, $I^T$ is some arbitrary intensity value at the focal spot of the lens, and $I^A$ is the actual intensity at that spot calculated via TMM. For this problem, we want to minimize the FOM over parameters $a,b,c$. We choose to optimize only over these parameters as we found very little dependence of the lens performance on the scatterer rotation. We initialize a grid of identical ellipsoidal scatterers with nominal semi-major axis radii $a=b=100nm$ and $c=300nm$, and a lattice periodicity of $450nm$. The radii are allowed to vary between $40 nm$ and $150 nm$ for axes a and b, and between $40nm$ and $300 nm$ for semi-major axis c. These parameters were picked in such a way that the scatterer phase responses span a $0-2\pi$ phase shift over this parameter range. The maximum radii of the semi-major axes are chosen in such a way that a circumscribing sphere with the radius of the largest semi-major axis radius of one particle cannot overlap with the surface of any neighboring particle. This is due to limitations in our method, as we compute the inter-scatterer coupling by assuming the incident electric field onto each particle is composed of the incident field and the field scattered from the surfaces of spheres inscribing the ellipsoidal particles \cite{t_matrix_ellip,EGEL2017103}. We also choose to cutoff our field expansions at the multiple order of $l$=3 \cite{Zhan:18}.

Fig. \ref{fig:inverse_fields} shows the performance of the designed lens. Fig. \ref{fig:inverse_fields} B shows that there is a clear focal spot at $10 \mu m$. All electric field in this work were calculated by using our extension of the CELES code \cite{EGEL2017103}. The efficiency of the lens was calculated by fitting a Gaussian shape to the field profile at the focal spot $z=10 \mu m$, for $x=0$ as shown in Fig. \ref{fig:inverse_fields} D. Then we found the full-with at half maximum of the Gaussian, and integrated the intensity of the field at that focal spot, and divided it over the total intensity of the incident light. This quantity is defined as the efficiency of the lens $\eta$ for some lens with focal length $F$ given by: 

\begin{equation}
\eta = \frac{\int\int_{\Omega} E^*(x,y, z = F) E(x,y, z = F) dx dy}{\int\int_{x,y} E^*(x,y,z=0 \mu m ) E(x,y,z=0) dx dy}
\end{equation}
$$\Omega := x^2 + y^2 < \left(3 \times FWHM\right)^2 $$
Here $\Omega$ is the surface around the focal spot which we integrate over. The efficiency calculated for this lens is 3.38\%. 

\subsection{Inverse design of Polarization switched focal length lens}

\begin{figure}[hbt!]
\begin{center}
\includegraphics[scale=0.65]{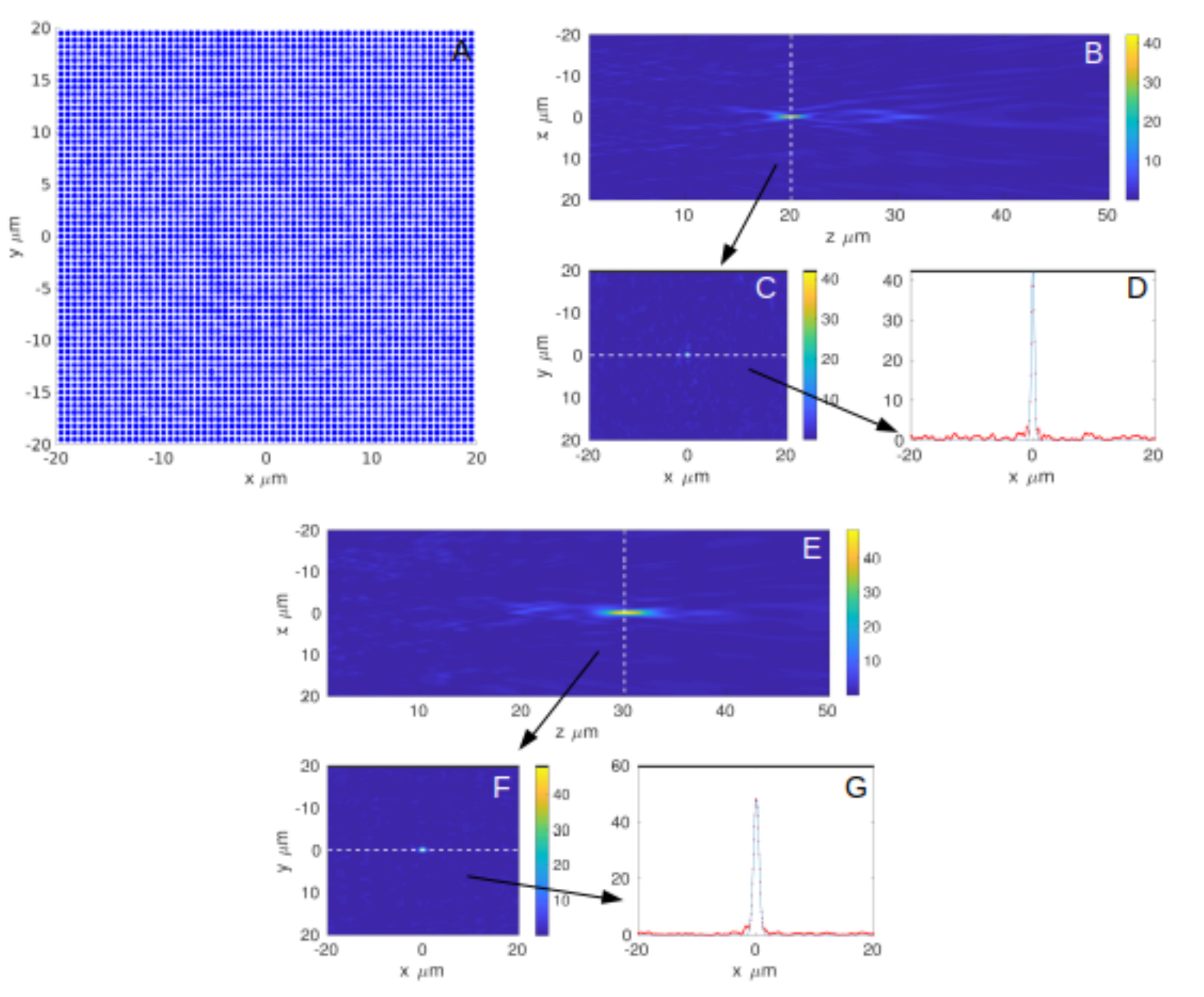}
\end{center}
\caption{\textbf{A} Scatterer distribution of the polarization multiplexed lens. Lattice periodicity is 650 nm, radii were limited to range from 40 nm to 292.5 nm for the $a$ and $b$ axes, and $0$ to $357.5 nm$ for the c axis. For the initial condition, all of the semi-major axis radii were set to $250 nm$, and the rotations were set to $0$ radians. In the final parameter distribution, the scatterers look very similar, and indeed, the minimum semi-major axis radius in the design is $\sim 205 nm$ and the maximum is $\sim 289 nm$. \textbf{B-D} Are field distributions correspond to x-polarized light, and \textbf{E-G} correspond to y-polarized light. \textbf{B,E} are scattered field slices in the x-z plane at $y=0\mu m$. \textbf{C,F} are x-y profiles at each focal spot. \textbf{C} is a slice at $z=20 \mu m$, and \textbf{E} is a slice at $z=30 \mu m$. \textbf{D,G} are Gaussian fits at each focal spot.}
\label{fig:multiplexed_device}
\end{figure}

Then we designed a lens with a diameter of 40 $\mu m$, and focal lengths of 20 $\mu m$ (NA $\sim 0.71$) and 30 $\mu m$ (NA $\sim 0.55$) for the x and y polarizations respectively. The lattice constant for this lens was taken to be $650 nm$. Semi-major axis radii $a$ and $b$ were allowed to range between $40 nm$ and $292.5 nm$. Semi-major axis radius $c$ was allowed to range between $40 nm$ and $357.5 nm$. The azimuthal rotation around the $z$ axis of the scatterers, was allowed to range from $-\pi/2$ to $\pi/2$.

The optimization problem was framed as a min-max optimization problem \cite{Pestourie:18}. For this optimization, we write the total FOM as a sum of FOM's for each polarization, given by
\begin{equation}
f = f_x + f_y
\end{equation}
with $f_x$ and $f_y$ being the figures of merit for the x and y polarizations respectively, and are \begin{equation}
f_x = (I_{max}^T(0,0, 20 \mu m) - I^A(0,0,20 \mu m)+I^T_{min}(0,0, 30 \mu m) - I^A(0,0,30 \mu m))^2
\end{equation}
\begin{equation}
f_y = (I_{max}^T(0,0, 30 \mu m) - I^A(0,0, 30 \mu m)+I_{min}^T(0,0, 20 \mu m) - I^A(0,0, 20 \mu m))^2
\end{equation}
Here, $I^T_{max}$ is some arbitrary large value (we chose 200), denoting the fact that light intensity at that spot should be maximized, while $I^T_{min}$ is a regularization term, denoting that the field intensity at that point should be kept small. To design this device, we minimize the maximum (worst) FOM iteratively until we converge to a local minimum:
\begin{equation}
\min_{P \in \{a,b,c,\phi\}} \max(f_x, f_y)
\end{equation}
The performance of the final device is shown in Fig. \ref{fig:multiplexed_device}. There is a clear focal spot at $z = 20 \mu m$ for x-polarized light, and no focal spot at $z=30 \mu m$ (Fig. \ref{fig:multiplexed_device}B) and for x-polarized light, we see a focal spot at $z=20\mu m$ and no focal spot at $z=30 \mu m$ (Fig. \ref{fig:multiplexed_device}E).

We calculated the efficiency of this lens for each polarization using the method described in the previous section, and found values of $\eta  = 2.31\%$ for the x-polarization and $\eta = 3.38 \%$ for y polarization. Another relevant quantities we can define to characterize the performance of this device are the contrast ratios of the focal spots. We define and report two different contrast quantities. The first one is the ratio between the value of intensity at the focal spot, where light should be maximized, to the ratio of light at the focal spot of the orthogonal polarization. We found the values for these ratios to be $\frac{I(0,0,30\mu m)}{I(0,0,20 \mu m} = 8.75$ for y polarized light and $\frac{I(0,0,20\mu m)}{I(0,0,30 \mu m} = 5.11$ for x polarized light. The second ratio we define to be the intensity at the focal spot for one polarization to the intensity at that same spot for the orthogonal polarization. We found these values to be $\frac{I_x(0,0,20\mu m}{I_y(0,0,20)} = 5.58$ and $\frac{I_y(0,0,30\mu m}{I_x(0,0,30)} = 5.92$.
\section{Metasurface lens optimization}
\begin{figure}[hbt!]
\begin{center}
\includegraphics[scale=0.4]{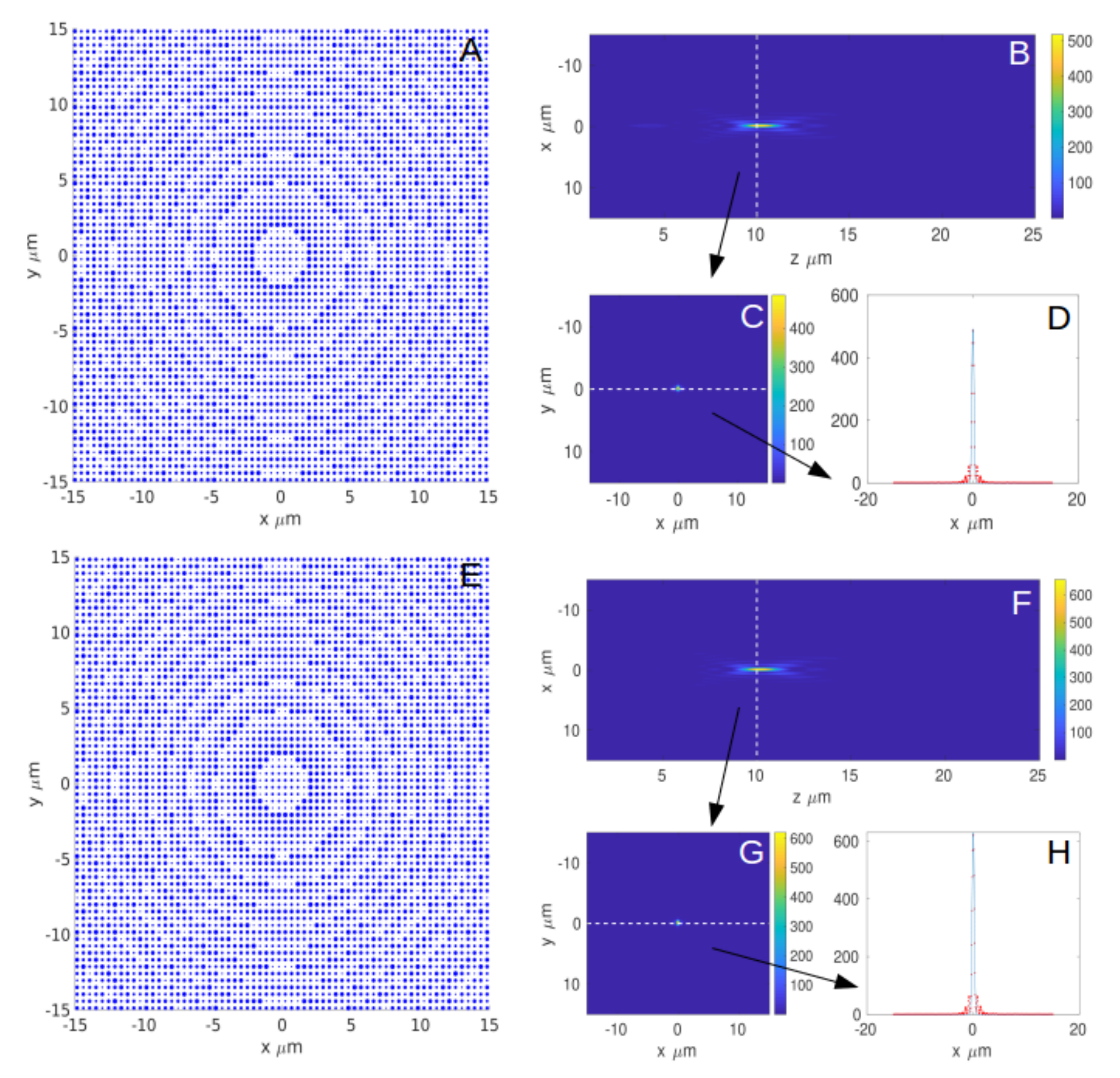}
\end{center}
\caption{Figs. \textbf{A-D} correspond to the forward designed lens, and Figs. \textbf{E-H} to the optimzied lens. \textbf{A,E} are the scatterer distributions. \textbf{E,F} are the x-z slices of the resulting field profile at $y=0\mu m$. \textbf{C,G} correspond to the x-y field slice at $z=10\mu m$. \textbf{D,H} are the Gaussians fitted to the field profiles at their focal spot with $y=0\mu m$. The forward design lens efficiency was determined to be $25.59\%$, and the optimzied efficiency was calculted to be $32.00\%$}
\label{fig:fwd_inv_dev}
\end{figure}

Finally, we discuss the optimization of a forward designed metalens using our inverse design method. By using rigorous coupled wave analysis (RCWA), 
we computed the phase and amplitude response of a library of ellipsoidal scatterers with periodic boundary conditions \cite{S4}. The ellipsoidal scatterers we chose for this design have identical geometric and material properties to those described in section 4.1. Then we discretized the design space in the x-y plane, using a scatterer periodicity of 450 nm, and by using the phase equation for a lens given by
\begin{equation}
\phi(x,y) = \frac{2\pi}{\lambda}(\sqrt{(x^2+y^2)+f^2}-f)
\end{equation}
we placed a scatterer at each discrete point $(x,y)$, with a phase response closest to the phase needed to focus light given by equation (21). The lens we designed has a diameter of 30$\mu m$ and a focal length of $10\mu m$. This devices performance is summarized in Fig. \ref{fig:fwd_inv_dev}. By using the same approach from the previous sections, we calculated the device's efficiency to be 25.59\%.

To optimize this device, we started off with the scatterer distribution given by the forward design as the initial condition and maximize the light intensity at a the focal spot. The performance of the optimized device is summarized in Fig \ref{fig:fwd_inv_dev}. The efficiency of this device was calculated to be 32.00\%, which is a 6.41\% improvement over the forward design lens. On average, each individual scatterer was changed by approximately $3.03 nm$ along the $a$ axis, $4.8 nm$ along the $b$ axis, and $0.17 nm$ along the c axis. The standard deviations for each axis are $3.53 nm$, $4.89 nm$, $0.42 nm$ respectively. The maximum changes for each axis were $33.42 nm$, $33.43 nm$ and $5.57 nm$ respectively. Its worth noting that this improvement implies that lenses designed by the conventional forward design methods are not necessarily globally optimal in the metasurface design space, even for high contrast designs. We can also see that the initial conditions are very important for the final design, as starting with identical ellipsoids, the final design provides very low efficiency. In fact, based on our analysis, we believe that our inverse design method will be more suitable for optimization type of problem, where the initial conditions are developed based on intuition and prior knowledge. 

\section{Discussion}
We demonstrated a new optimization method for designing large area dielectric metasurfaces made of ellipsoidal scatterers based on the adjoint method and a generalization of GMMT. Starting from an array of identical ellipsoidal scatterers, we designed a high NA ($\sim 0.83$) lens and a polarization multiplexed lens that focuses light at $30\mu m$ and $20 \mu m$ based on the polarization of incident light. We have also shown that starting with a forward-designed lens as an initial condition, a higher efficiency design can be obtained via optimization.

We note that all the reported devices were designed at refractive index $n = 3.56$. As our method requires the bounding spheres of ellipsoidal scatterers not to overlap other ellipsoidal scatterers, we are limited by the aspect ratio and density of the ellipsoids, and only with high index ellipsoids we can maintain low density of scatterers while spanning the whole $0-2\pi$ phase. Unfortunately, there is currently no straightforward way to fabricate these structures with such a high index. One solution could be to use a high index resin in additive manufacturing \cite{HighIndex3D}. It is also possible to fabricate cylindrical scatterers at high refractive indicies by using traditional lithography. This would require a further generalization of the T-Matrix method to expand the incident and scattered fields in terms of spheroidal wave-functions instead of SVWF or by using the plane wave coupling method \cite{plane-wave-coupling}. 

\section*{Funding}
The research is funded by Samsung GRO and NSF grant number 1825308.

\section*{Disclosures}
``The authors declare that there are no conflicts of interest related to this article.''






\section*{Appendix A: T-Matrix derivatives}
The T-Matrix of the ellipsoid requires the computation of the $Q$ and $RgQ$ matrices that represent the coupling between the scattered field and the incident field to the surface fields respectively. It depends only on the geometric and material properties of particle itself, and the wavelength of excitation. The T-Matrix is then given by\cite{waterman_ebcm,mishchenko_book,doicu_book}:
\begin{equation}
T = RgQ (Q)^{-1}
\end{equation}
The $Q$ matrix is square, and is composed of four square submatrices $\bar{P}$, $\bar{R}$, $\bar{S}$, and $\bar{U}$, given by:
\begin{equation}
Q = \begin{bmatrix}
\bar{P} & \bar{R} \\
\bar{S} & \bar{U} \end{bmatrix},
\end{equation}
These individual square matrices are given by \cite{t_matrix_ellip}:
\begin{equation}
\bar{P}_{lml'm'} = -i k k_s J^{(21)}_{lml'm'} - i k^2 J^{(12)}_{lml'm'},
\end{equation}
\begin{equation}
\bar{R}_{lml'm'} = -i k k_s J^{(11)}_{lml'm'} - i k^2 J^{(22)}_{lml'm'},
\end{equation}
\begin{equation}
\bar{S}_{lml'm'} = -i k k_s J^{(22)}_{lml'm'} - i k^2 J^{(11)}_{lml'm'},
\end{equation}
\begin{equation}
\bar{U}_{lml'm'} = -i k k_s J^{(12)}_{lml'm'} - i k^2 J^{(21)}_{lml'm'},
\end{equation}
where the $J$ terms represent integrals over the surface of the particle, and are given by:
\begin{equation}
J^{(pq)}_{lml'm'} = (-1)^m \int_S dS \hat{n}({\bf r}) \cdot {\bf \Psi}_{p,l',m'}^{(1)} (k_s r, \theta,\phi) \times {\bf \Psi}_{q,l,-m}^{(3)}(kr,\theta,\phi),
\end{equation}
where $S$ is the surface bounding the particle, $dS$ is infinitesimal surface area, and $\hat{n}$ is a outward pointing unit normal at $dS$. The SVWFs ${\bf\Psi}^{(1)}$ and $\bf{\Psi}^{(3)}$ are given by \cite{mishchenko_book}:
\begin{equation}
{\bf \Psi}^{(\nu)}_{1lm}({\bf r}) = \frac{e^{i m \phi}}{\sqrt{2l(l+1)}}b_l(kr)\big[im\pi_{lm}(\theta)\hat{\theta}-\tau_{lm}(\theta)\hat{\phi}\big],
\end{equation}
\begin{equation}
\begin{aligned}
{\bf \Psi}^{(\nu)}_{2lm}({\bf r}) = &\frac{e^{i m \phi}}{\sqrt{2l(l+1)}}\bigg\{l(l+1)\frac{b_l(kr)}{kr}P^{|m|}_l(cos\theta)\hat{r}
\\ &+\frac{1}{kr}\frac{\partial (kr b_l(kr))}{\partial (kr)}\big[\tau_{lm}(\theta)\hat{\theta}+im\pi_{lm}(\theta)\hat{\phi}\big]\bigg\},
\end{aligned}
\end{equation}
Here we have defined:
\begin{equation}
\pi_{lm}(\theta) = \frac{P^{|m|}_l(cos\theta)}{sin\theta},
\tau_{lm}(\theta) = \frac{\partial P^{|m|}_l(cos\theta)}{\partial \theta}.
\end{equation}
$P^m_l(x)$ is the associated Legendre polynomial. $j_l$ is the spherical Bessel function of order $l$, and $b_l$ is either a spherical Bessel function ($j_l$) for $\nu = 1$ or spherical Hankel function of the first kind ($h^{(1)}_l$) of order $l$ for $\nu = 3$, depending on whether $RgQ$ or $Q$ is being computed.
In spherical coordinates, the product of the unit normal and the infinitesmal area is:
\begin{equation}
dS \hat{n}({\bf r}) = r^2 sin(\theta){\bf \sigma}({\bf r}) d\theta d\phi,
\end{equation}
and ${\bf \sigma}$ is given by:
\begin{equation}
{\bf \sigma}({\bf r}) = \hat{r} - \hat{\theta}\frac{1}{r}\frac{\partial r}{\partial \theta} - \hat{\phi}\frac{1}{r sin\theta}\frac{\partial r}{\partial \theta}.
\end{equation}
In this case, $r$ is parameterizing the surface of a particle, and for an ellipsoid in spherical coordinates, $r$ is given by:
\begin{equation}
r(\theta,\phi) = \left[ sin^2\theta \left(\frac{cos^2\phi}{a^2}+\frac{sin^2\phi}{b^2}\right) + \frac{cos^2\theta}{c^2}\right]^{-1/2}
\end{equation}
To compute $RgQ$ rather than $Q$, we simply need to replace ${\bf \Psi}^{(3)}$ in the $J$ integrals with ${\bf \Psi}^{(1)}$.

The derivative of the T-Matrix of a particle with respect to some parameter $p$ is given by:
\begin{equation}
\frac{\partial T}{\partial p} = \left(\frac{\partial RgQ}{\partial p} - T \frac{\partial Q}{\partial p}\right)Q^{-1},
\end{equation}
Hence we need to find the derivatives of the sub-matrices $\bar{P}$, $\bar{R}$, $\bar{S}$, and $\bar{U}$ with respect to $p$. This requires us to take the derivatives of the surface integrals $J$ with respect to the parameter $p$. In general, our parameter of interest $p$ will be some geometric quantity that determines the shape of the surface of integration $S$. In the specific case of ellipsoidal scatterers, they will be the three independent axes $a$, $b$, and $c$ along the $x$, $y$, and $z$ axes respectively.

The expressions for the derivatives with respect to a spatial variable ($a$, $b$, $c$) are as follows where $p$ represents any of the ellipsoid axes:
\begin{equation}
\begin{aligned}
\frac{\partial J^{(11)}_{lml'm'}}{\partial p} = & -i  \iint \alpha_{lml'm'}  \left( m'\pi_{l'm'} \tau_{lm}+m \pi_{lm} \tau_{l'm'}\right)
\\ & \left[r\left(k\frac{\partial b_l}{\partial p} j_{l'} + k_s b_l \frac{\partial j_{l'}}{\partial p}\right)+2b_l j_{l'}\right]r sin\theta d\theta d\phi,
\end{aligned}
\end{equation}
\begin{equation}
\begin{aligned}
\frac{\partial J^{(12)}_{lml'm'}}{\partial p}  = &  \iint \alpha_{lml'm'} \bigg\{\bigg[ \frac{\partial R^{(12)}_{lml'm'}}{\partial r}+(\Theta^{(12)}_{lml'm'} E_\theta+\Phi^{(12)}_{lml'm'} E_\phi)\frac{\partial \rho_{l,l'}}{\partial r}\bigg] \frac{\partial r}{\partial p} 
\\ &+\bigg( \Theta^{(12)}_{lml'm'} \frac{\partial E_\theta}{\partial p}+\Phi^{(12)}_{lml'm'}\frac{\partial E_\phi}{\partial p}\bigg) \rho_{l,l'}\bigg\}d\theta d\phi,
\end{aligned}
\end{equation}
\begin{equation}
\begin{aligned}
\frac{\partial J^{(21)}_{lml'm'}}{\partial p}  = &  \iint \alpha_{lml'm'} \bigg\{\bigg[ \frac{\partial R^{(21)}_{lml'm'}}{\partial r}+(\Theta^{(21)}_{lml'm'} E_\theta+\Phi^{(21)}_{lml'm'} E_\phi)\frac{\partial \rho_{l,l'}}{\partial r}\bigg] \frac{\partial r}{\partial p}
\\ &+\bigg( \Theta^{(21)}_{lml'm'} \frac{\partial E_\theta}{\partial p}+\Phi^{(21)}_{lml'm'}\frac{\partial E_\phi}{\partial p}\bigg) \rho_{l,l'}\bigg\}d\theta d\phi,
\end{aligned}
\end{equation}
\begin{equation}
\begin{aligned}
\frac{\partial J^{(22)}_{lml'm'}}{\partial p}  = & \iint \alpha_{lml'm'} \bigg\{\bigg[\frac{\partial R^{(22)}_{lml'm'}}{\partial r}+\frac{\partial \Theta^{(22)}_{lml'm'}}{\partial r}E_\theta+\frac{\partial \Phi^{(22)}_{lml'm'}}{\partial r}E_\phi\bigg]\frac{\partial r}{\partial p} 
\\ & +\Theta^{(22)}_{lml'm'} \frac{\partial E_\theta}{\partial p}+\Phi^{(22)}_{lml'm'} \frac{\partial E_\phi}{\partial p} \bigg\} d\theta d\phi,
\end{aligned}
\end{equation}
where we have defined:
\begin{equation}
\alpha_{lml'm'} = \frac{(-1)^m(1+(-1)^{m'-m})(1+(-1)^{l'+l+1}}{2\sqrt{l(l+1)l'(l'+1)}}e^{i(m'-m)\phi}
\end{equation}
$k$ and $k_s$ are the k vectors of light in the medium surrounding the particle, and in the particle itself. Then we define:
\begin{equation}
E_\theta = \frac{cos^2\phi}{a^2} + \frac{sin^2\phi}{b^2} - \frac{1}{c^2}
\end{equation}
\begin{equation}
E_\phi = \frac{1}{b^2}-\frac{1}{a^2},
\end{equation}
\begin{equation}
\rho_{l,l'} = r^3 j_{l'}b_l,
\end{equation}
Now, we can define the specific terms used to construct each $J$ surface integral. For $J^{(12)}$, we define:
\begin{equation}
\begin{aligned}
 \frac{\partial R^{(12)}_{lml'm'}}{\partial r} = \frac{sin\theta}{k} (mm' \pi_{l'm'}\pi_{lm}+\tau_{l'm'}\tau_{lm})\bigg(j_{l'}\frac{\partial (kr b_l)}{\partial (kr)}\\+r\bigg(k_s \frac{\partial j_{l'}}{\partial r}\frac{\partial (kr b_l)}{\partial (kr)}+k j_{l'} \frac{\partial}{\partial r}\left(\frac{\partial (kr b_l)}{\partial (kr)}\right)\bigg)\bigg),
\end{aligned}
\end{equation}

\begin{equation}
\Theta^{(12)}_{lml'm'} = -\frac{sin\theta}{k}l(l+1)P^{|m|}_l\tau_{l'm'},
\end{equation}

\begin{equation}
\Phi^{(12)}_{lml'm'} = -i\frac{sin\theta}{k} l(l+1)m'P^{|m|}_l\pi_{l'm'}.
\end{equation}
For $J^{(21)}$, we define:
\begin{equation}
\begin{aligned}
 \frac{\partial R^{(21)}_{lml'm'}}{\partial r} = - \frac{sin\theta}{k_s} (mm' \pi_{l'm'}\pi_{lm}+\tau_{l'm'}\tau_{lm})\bigg(\frac{\partial (k_s r j_{l'})}{\partial(k_s r)}b_l
\\+r\bigg(k_s \frac{\partial}{\partial r}\left(\frac{\partial (k_s r j_{l'})}{\partial (k_s r)}\right)b_l + k \frac{\partial b_l}{\partial r}\frac{\partial (k_s r j_{l'})}{\partial(k_s r)}\bigg)\bigg),
\end{aligned}
\end{equation}
\begin{equation}
\Theta^{(21)}_{lml'm'} = \frac{sin\theta}{k_s}l'(l'+1)P^{|m'|}_{l'}\tau_{lm},
\end{equation}

\begin{equation}
\Phi^{(21)}_{lml'm'} = -i\frac{sin\theta}{k_s} l'(l'+1)mP^{|m'|}_{l'}\pi_{lm}.
\end{equation}
Finally, for $J^{(22)}$ we define:
\begin{equation}
\begin{aligned}
\Theta^{(22)}_{lml'm'}= & i \frac{r^2 sin\theta}{k k_s} \bigg(m'l(l+1) \frac{\partial (k_s r j_{l'})}{\partial (k_s r)} b_l P^{|m|}_l \pi_{l'm'}
\\ & +ml'(l'+1)j_{l'}\frac{\partial (kr b_l)}{\partial (kr)}P^{|m'|}_{l'}\pi_{lm}\bigg)
\end{aligned}
\end{equation}
\begin{equation}
\begin{aligned}
\Phi^{(22)}_{lml'm'} = & \frac{r^2 sin\theta}{k k_s}\bigg(l'(l'+1)j_{l'}P^{|m'|}_{l'}\frac{\partial (kr b_l)}{\partial (kr)}\tau_{lm}
\\ & -l(l+1)\frac{\partial (k_s r j_{l'})}{\partial (k_s r)}\tau_{l'm'}b_lP_{lm}\bigg)
\end{aligned}
\end{equation}
and the three derivative terms:
\begin{equation}
\begin{aligned}
\frac{\partial R^{(22)}_{lml'm'}}{\partial r} = &-i\frac{sin\theta}{k k_s}( m'\pi_{l'm'} \tau_{lm}+m \pi_{lm} \tau_{l'm'})
\\ & \bigg(k \frac{\partial}{\partial r}\left(\frac{\partial (kr b_l)}{\partial (kr)}\right)\frac{\partial (k_s r j_{l'})}{\partial (k_s r)}
\\ &+ k_s\frac{\partial}{\partial r} \left(\frac{\partial (k_s r j_{l'})}{\partial (k_s r)}\right)\frac{\partial (kr b_l)}{\partial (kr)}\bigg)
\end{aligned}
\end{equation}
\begin{equation}
\begin{aligned}
\frac{\partial \Theta^{(22)}_{lml'm'}}{\partial r} = & i\frac{sin\theta}{k k_s}\bigg(ml'(l'+1)P^{|m'|}_{l'}\pi_{lm}\bigg(2r \frac{\partial (kr b_l)}{\partial (kr)} j_{l'}
\\ &+r^2\bigg(k \frac{\partial}{\partial r}\left(\frac{\partial (kr b_l)}{\partial (kr)}\right) j_{l'}+k_s \frac{\partial j_{l'}}{\partial r} \frac{\partial (kr b_l)}{\partial (kr)}\bigg)\bigg)
\\&+m'l(l+1)P^{|m|}_l\tau{l'm'}\bigg(2rb_l\frac{\partial (k_s r j_{l'})}{\partial (k_s r)} 
\\&+r^2\bigg(k \frac{\partial b_l}{\partial r}\frac{\partial (k_s r j_{l'})}{\partial (k_s r)}+k_sb_l\frac{\partial}{\partial r} \left(\frac{\partial (k_s r j_{l'})}{\partial (k_s r)}\right)\bigg)\bigg)\bigg)
\end{aligned}
\end{equation}
\begin{equation}
\begin{aligned}
\frac{\partial \Phi^{(22)}_{lml'm'}}{\partial r} = & \frac{sin\theta}{k k_s} \bigg(l'(l'+1)P^{|m'|}_{l'}\tau_{lm}\bigg(2r\frac{\partial (kr b_l)}{\partial (kr)}j_{l'}
\\&+r^2\bigg( \frac{\partial}{\partial r}\left(\frac{\partial (kr b_l)}{\partial (kr)}\right)j_{l'}+k_s \frac{\partial j_{l'}}{\partial r}\frac{\partial (kr b_l)}{\partial (kr)}\bigg)\bigg)
\\&-l(l+1)P^{|m|}_l\tau_{l'm'}\bigg(2r b_l \frac{\partial (k_s r j_{l'})}{\partial (k_s r)} 
\\&+r^2\bigg(k\frac{\partial b_l}{\partial r}\frac{\partial (k_s r j_{l'})}{\partial (k_s r)}\frac{\partial}{\partial r} \left(\frac{\partial (k_s r j_{l'})}{\partial (k_s r)}\right)\bigg)\bigg)\bigg).
\end{aligned}
\end{equation}
Now with these $J$ integrals, we can compute the quantity $\frac{\partial T}{\partial p}$ for a given axis of an ellipsoid in its own particle frame where $a$, $b$, and $c$ are aligned along the $x_{part}$, $y_{part}$, and $z_{part}$ axes. 

In addition to computing the response of the T-Matrix of the ellipsoid to the contraction or extension of one of its axes, we are also interested in its response to rotations about the $z_{part}$ axis. To do this we will first define the transformation of the T-Matrix or a derivative matrix from the particle frame to some rotated lab frame that has new axes $x_{lab}$ and $y_{lab}$, but shares $z_{lab} = z_{part}$. Given some rotation angle $\phi_{rot}$, we can then define our new axes:

\begin{subequations}
\begin{align}
x_{lab} =& x_{part}cos(\phi_{rot})+y_{part}sin(\phi_{rot}) \\
y_{lab} =& -x_{part}sin(\phi_{rot})+y_{part}cos(\phi_{rot}) \\
z_{lab} =& z_{part}
\end{align}
\end{subequations}
The general form of this orthogonal transformation in three dimensions can be represented by the Euler angles $\alpha$, $\beta$, and $\gamma$. The general transformation of an element of an operator $O$ from the particle frame to the lab frame can be written as\cite{mishchenko_book}:
\begin{equation}
O^{lab}_{plmp'l'm'}(\alpha,\beta,\gamma) = \sum^l_{m_1 = -l}\sum^{l'}_{m_2 = -l'}D^{l}_{mm_1}(\alpha,\beta,\gamma)O^{particle}_{plm_1p'l'm_2}D^{l'}_{m_2m'}(-\gamma,-\beta,-\alpha),
\end{equation}
where the $D$ operator is a Wigner D-function. It can be represented as:
\begin{equation}
D^l_{m'm}(\alpha,\beta,\gamma) = e^{-im'\alpha}d^l_{m'm}(\beta)e^{-im\gamma},
\end{equation}
where $d^l_{m'm}(\beta)$ is Wigner's (small) d-matrix given by:
\begin{equation}
d^l_{m'm}(\beta) = \bra{l,m'}e^{-i\beta J_y}\ket{l,m}.
\end{equation}
However, as we are only concerned with rotations about the $z$ axis, we can simplify our expressions knowing that $\alpha$ is our only nonzero angle, and equation (56) becomes:
\begin{equation}
O^{lab}_{plmp'l'm'}(\alpha,0,0) = \sum^l_{m_1 = -l}\sum^{l'}_{m_2 = -l'}D^{l}_{mm_1}(\alpha,0,0)O^{particle}_{plm_1,p'l'm_2}D^{l'}_{m_2m'}(0,0,-\alpha).
\end{equation}
In this case, our $D$ operator has a much simplified form:
\begin{subequations}
\begin{align}
D^{l}_{m'm}(\alpha,0,0) = & e^{-im'\alpha}\delta_{m'm} \\
D^{l}_{m'm}(0,0,\gamma) = & e^{-im\gamma}\delta_{m'm}.
\end{align}
\end{subequations}
Combining equations (46), (47a), and (47b), we obtain a simple expression transforming $O$ from the particle frame to the lab frame:
\begin{equation}
O^{lab}_{plmp'l'm'}(\alpha) = e^{i(m'-m)\alpha}O^{particle}_{plmp'l'm'}.
\end{equation}

Equation (61) is applicable to for transforming both T-matrices and the derivative matrices computed in the particle frame into the lab frame. It also gives us a prescription for computing the derivative matrix with respect to the particle's angular orientation. We already have derivatives characterizing the response of the particle to contractions and extensions of its principal axes, and can now rotate these to a lab frame where the particle has an arbitrary angular orientation relative to the $z$ axis. We can now compute the derivative with respect to the particle's angular orientation $\alpha$ as:
\begin{equation}
\frac{\partial T^{lab}_{plmp'l'm'}(\alpha)}{\partial \alpha} = i(m'-m)e^{i(m'-m)\alpha}T^{particle}_{plmp'l'm'}.
\end{equation}
With equations (31) and (58), we have characterized the derivatives of the T-Matrix representing an ellipsoid with respect to its axes and orientation.

These integrals are implemented in MATLAB, and performed using Gaussian quadrature. 

\section*{Appendix B: Scatterer Electromagnetic Response}

\begin{figure}[hbt!]
\begin{center}
\includegraphics[scale=0.4]{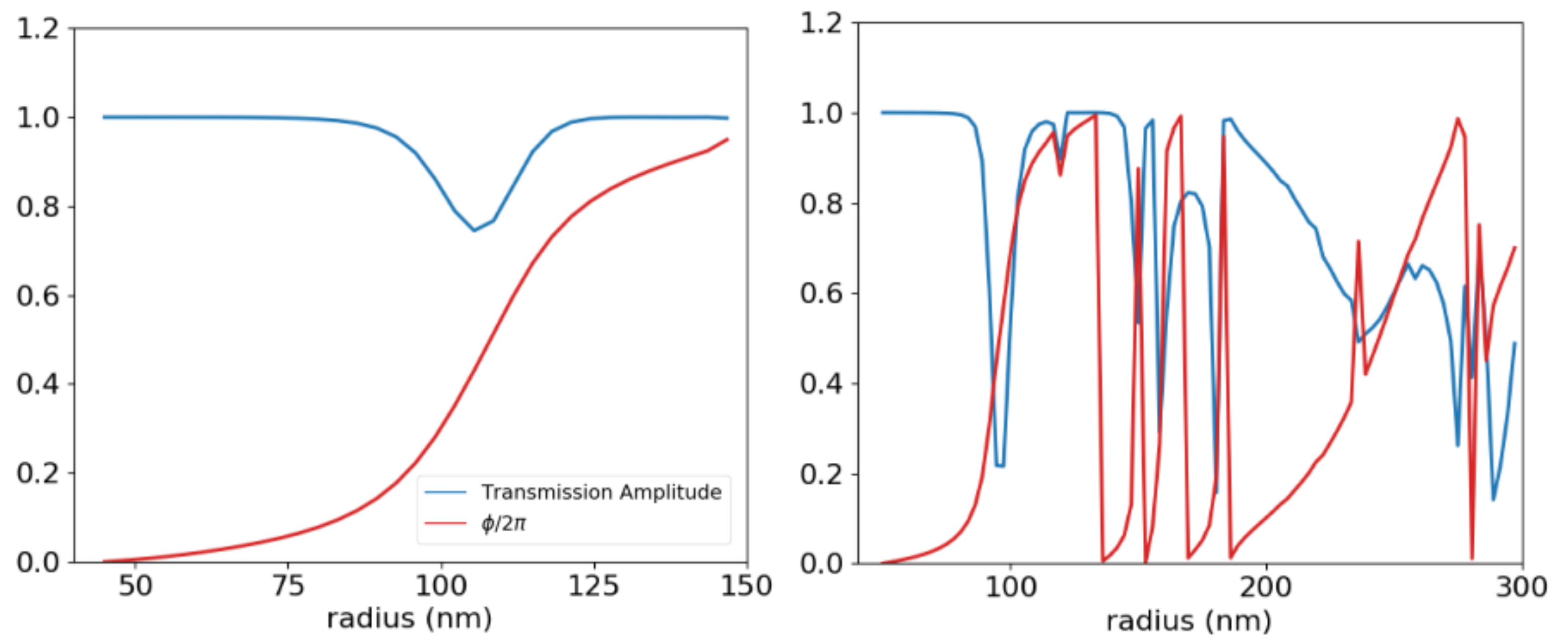}
\end{center}
\caption{Transmission of individual scatterers with periodic boundary conditions as a function of the radius of the ellipsoids (semi-major axes $a = b$). \textbf{A} is the plot of the complex transmission of the ellipsoids used from section 4.1 and 5. Ellipsoid height is fixed to be $600 nm$. The lattice constant is $450 nm$. \textbf{B} transmission response for ellipsoids outlined in section 4.2.  Ellipsoidal height is fixed at $715nm$ with a lattice constant of $650nm$.}
\label{fig:rcwa}
\end{figure}

\section*{Appendix C: Machine Specifications}
Ubuntu 16.04\\
MATLAB v9.5.0 R2018b with parallel computing toolbox v2.4
2x Intel E5-2620 at 2.1 GHz\\
NVIDIA Tesla K40 12 GB Memory running CUDA 9.1\\
64 GB DDR3 Memory\\
Our inverse and forward design methods are solved using a modified version of CELES. More details about celes are available from Egel et. al. \cite{EGEL2017103}.\\
CELES is available free of charge, and our implementation of the optimization algorithm is available upon request.

\end{document}